\documentclass[3p,times,procedia]{elsarticle}
\usepackage{nupha_ecrc}

\volume{00}

\firstpage{1}

\journalname{Nuclear Physics A}

\runauth{K. Nagashima}

\jid{nupha}

\jnltitlelogo{Nuclear Physics A}

\usepackage{amssymb}

\usepackage{lineno}

\usepackage[figuresright]{rotating}
\usepackage{xspace}

\newcommand{\gev}{\mbox{GeV}\xspace}
\newcommand{\gevc}{\mbox{GeV/$c~$}\xspace}

\newcommand{\mum}{\mbox{$\mu$m}\xspace}

\newcommand{\sqsn}{\mbox{$\sqrt{s_{_{NN}}}$}\xspace}
\newcommand{\pp}{\mbox{$p$+$p$}\xspace}

\newcommand{\auau}{\mbox{Au+Au}\xspace}

\newcommand{\raa}{\mbox{$R_{\rm AA}$}\xspace}
\newcommand{\pt}{\mbox{$p_{\rm T}$}\xspace}
\newcommand{\btoe}{\mbox{$\rm{b}{\rightarrow}\rm{e}$}\xspace}
\newcommand{\ctoe}{\mbox{$\rm{c}{\rightarrow}\rm{e}$}\xspace}

\newcommand{\dcat}{\mbox{$DCA_{T}$}\xspace}

\begin{document}

\begin{frontmatter}



\dochead{}

\title{PHENIX measurements of single electrons from charm and bottom decays at midrapidity in \auau collisions}


\author[label1]{K. Nagashima (for the PHENIX Collaboration)}\footnote{For the full PHENIX Collaboration author list and acknowledgments,
see Appendix ``Collaboration'' of this volume}
\address{Hiroshima University, Kagamiyama, Higashi-Hiroshima 739-8526 ,Japan}


\begin{abstract}
 Heavy quarks (charm and bottom) are a sensitive probe of the Quark-Gluon Plasma (QGP) created in high-energy nuclear collisions. The modification of their phase-space distribution in the QGP reflects the physical property of the QGP because they are generated in the early stage of collisions and subsequently propagate through the QGP.  PHENIX has published the first measurement of the nuclear modification factors \raa of separated charm and bottom hadron decay electrons in minimum bias Au+Au collisions at \sqsn $=$ 200 \gev using the 2011 data set. In these proceedings, we present $\raa({\ctoe})$ and $\raa({\btoe})$ in most central Au+Au collisions at \sqsn $=$ 200 GeV using 1/8 of our large 2014-2016 data set.
\end{abstract}

\begin{keyword} Quark-gluon plasma, PHENIX, Open heavy flavor, Energy loss of charm and bottom quarks


\end{keyword}

\end{frontmatter}


\section{Introduction}
 The PHENIX collaboration at the Relativistic Heavy Ion Collider has measured strong suppression of electrons from heavy flavor hadron decays in the Quark-Gluon Plasma (QGP) \cite{cite1,cite2}. This discovery opened the window of a heavy flavor probe in QGP. The energy loss of heavy quarks is a sensitive probe of the transport properties of the QGP, the diffusion coefficient $\it{D} \propto \eta/(\it{sT})$. Since bottom and charm quark masses are larger than $\Lambda_{\rm{QCD}}$ and the QGP temperature, they are produced by a hard scattering in the earliest stages of heavy-ion collisions and they experience the full-time evolution of QGP since flavor is conserved. Therefore, a modification of the heavy-flavor phase-space distribution in the QGP reflects the physical properties of the QGP. Furthermore, theoretical calculations indicate a mass ordering of quark energy loss in the QGP, ${\Delta}E_{\rm{light}} > {\Delta}E_{\rm{c}} > {\Delta}E_{\rm{b}}$, because both collisional and radiative energy loss are suppressed with increasing quark mass and the small-angle gluon radiation of heavy quarks is suppressed via the Dead-Cone effect \cite{cite5}. A comparison of the measured nuclear modification factor \raa of separated charm and bottom hadron decay electrons with pQCD NLO calculations can provide an understanding of the quark energy loss mechanism in the QGP and the diffusion coefficient of the QGP.\par
 PHENIX has published the first measurement of invariant yields and nuclear modification factors \raa of separated charm and bottom hadron decay electrons in minimum bias Au+Au at \sqsn = 200 \gev using a Bayesian unfolding technique applied simultaneously to the yield and the distance of closest approach in the transverse plane (\dcat) \cite{cite3}. In these proceedings, we present $R_{AA}(\ctoe)$ and $R_{AA}(\btoe)$ in 0-10\% central Au+Au collisions at \sqsn $=$ 200 \gev using 1/8 of our large 2014-2016 data set.

\section{Analysis}
 The silicon vertex detector (VTX) was installed at PHENIX in 2011 to measure precise displaced vertices. The VTX allows a separation of electrons from charm and bottom semi-leptonic weak decays each other and from backgrounds.  The distribution of \dcat of the track to the primary vertex measured with the VTX is a useful tool to separate charm and bottom hadron decay electrons because the lifetime of bottom hadrons (${c\tau_{B^{0}}}$ = 455 \mum) is longer than that of charm hadrons (${c\tau_{D^{0}}}$ = 123 \mum) and decay kinematics are different. In 2014-2016, PHENIX collected about 40 billion events in Au+Au at \sqsn = 200 \gev, which is 20 times larger than the 2011 data set.  This large data set will allow for the measurement of the centrality dependence of $\raa(\ctoe)$ and $\raa(\btoe)$.\par

\begin{figure}[htbp]
 \begin{minipage}{0.5\hsize}
  \begin{center}
   \includegraphics[width=1.0\columnwidth]{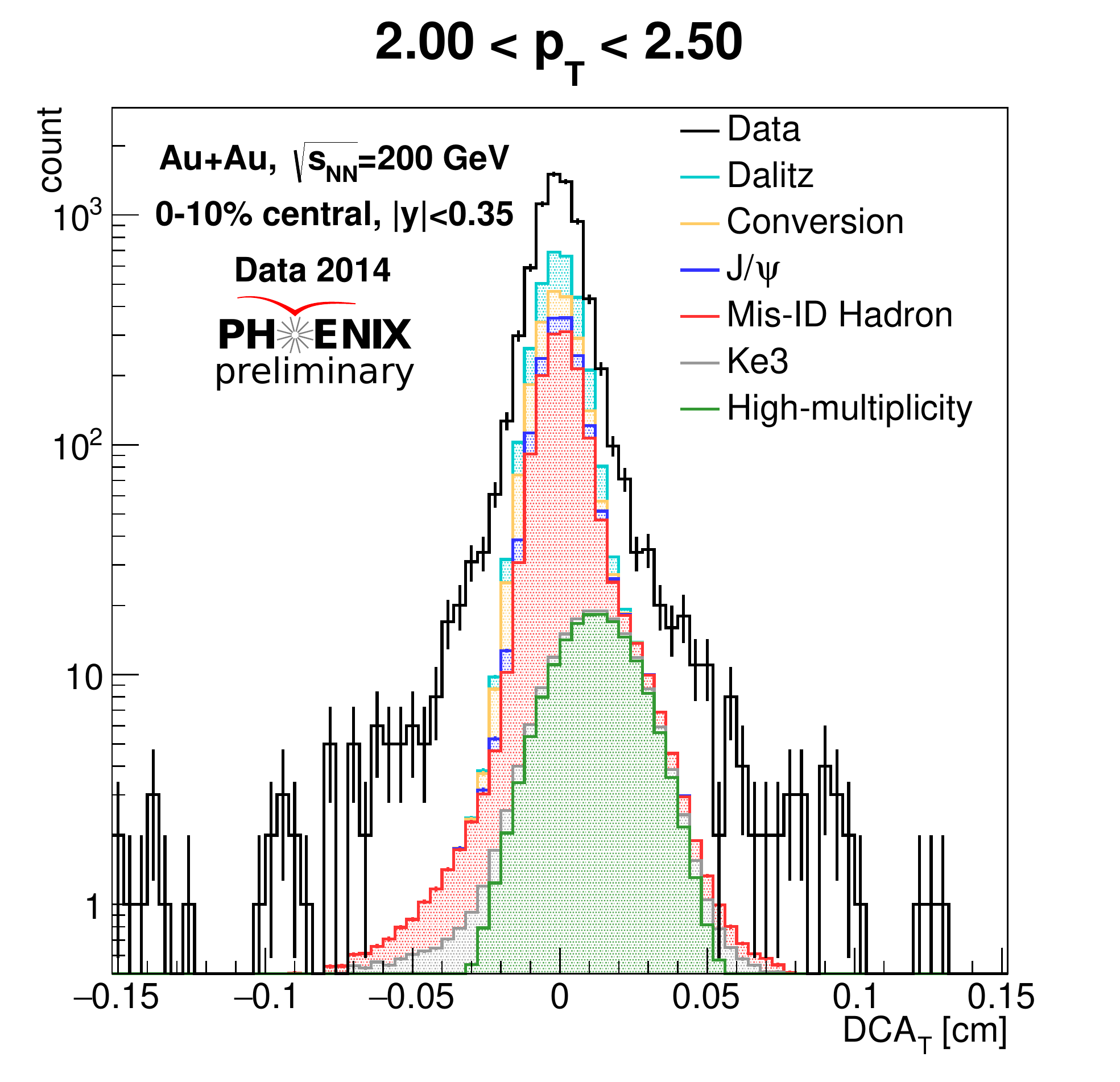}
  \end{center}
 \end{minipage}
 \begin{minipage}{0.5\hsize}
  \begin{center}
   \includegraphics[width=1.0\columnwidth]{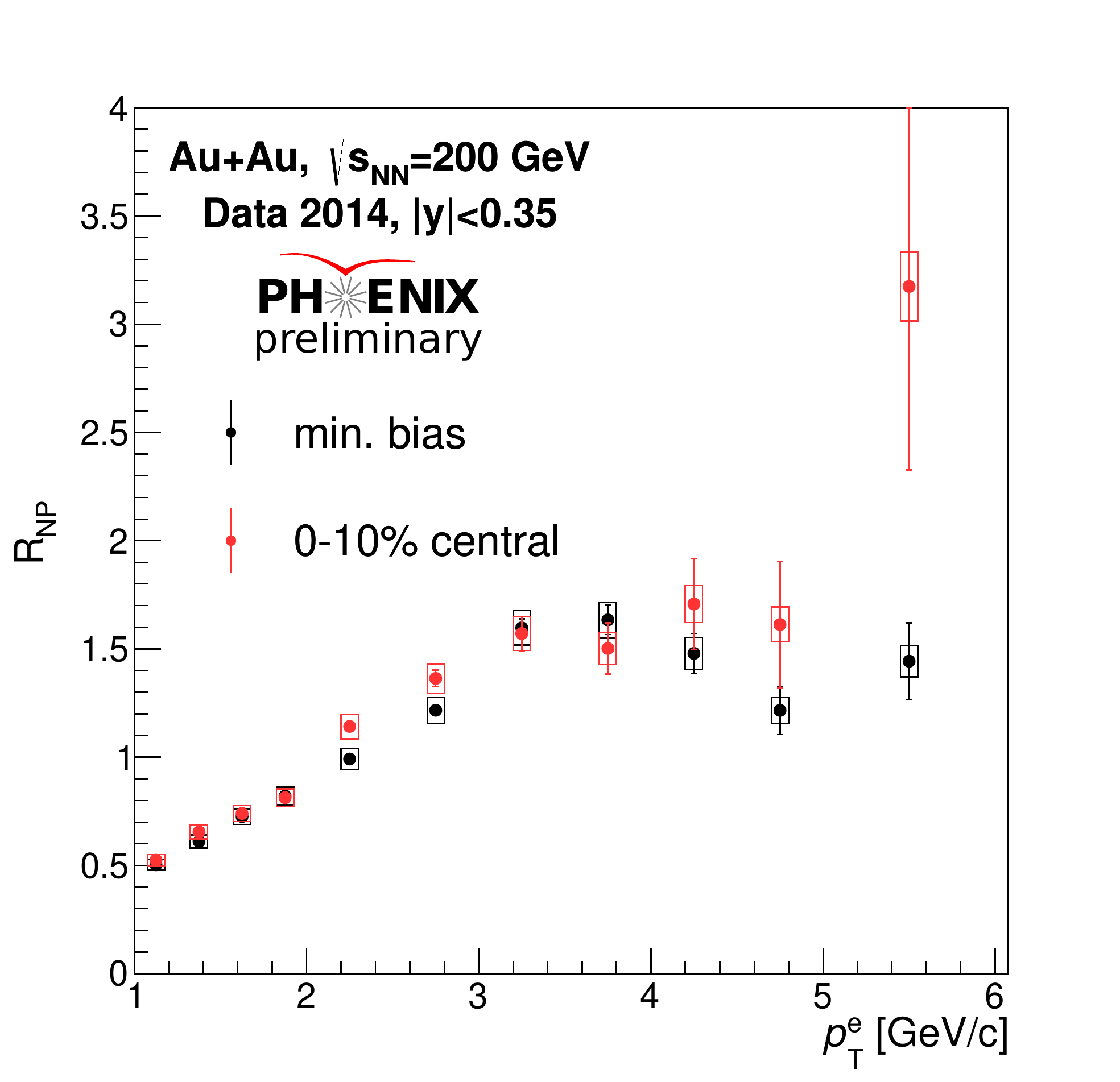}
  \end{center}
 \end{minipage}
 \caption{(left) The \dcat distribution for measured electrons (black histogram) and all background contributions (stacked) in the range $2.0 < \pt [\gevc] < 2.5$. (right) The non-photonic to photonic ratio, $R_{\rm{NP}}$, for minimum bias (black) and 0-10\% central (red) Au+Au collisions at \sqsn $=$ 200 \gevc. \label{fig:1}}
\end{figure}
 
  The \dcat of electron tracks is measured with the PHENIX central arm detector for 1.25 $<$ \pt $<$ 6.0 \gevc in 6 \pt bins. The background of the \dcat is composed of 3 contributions: photonic electrons from Dalitz decay and photon conversion, misidentified electrons due to random associations in a high-multiplicity environment, and non-photonic electrons from $J/\psi$ and kaon weak decay ($K_{e3}$)). Background shapes of photonic electrons and non-photonic electrons in \dcat distributions are estimated by the Cocktail method \cite{cite1}.  Mis-reconstructed electron tracks are caused by association between Drift Chamber tracks and random hits at VTX or random electron-ID hits at RICH in the high-multiplicity environment. This background \dcat distribution can be reproduced by a track-swapping method \cite{cite3} such as the event-mixing method.  Each component of the background \dcat distribution is shown in Fig.\ \ref{fig:1} (left).\par
  The largest contribution of background is from photonic electrons, which can be identified with an isolation cut at VTX because to the very close pair creation from internal and external photon conversions \cite{cite3}. On the other hand, non-photonic electrons including charm and bottom electrons are not removed by an isolation cut because of the larger opening angle. Therefore, the photonic to non-photonic electron ratio $R_{\rm{NP}}$ is calculated with inclusive electron samples and isolated electron samples to normalize photonic \dcat backgrounds shown in Fig.\ \ref{fig:1} (right).\par
  In this analysis, we apply the Bayesian inference technique simultaneously to the invariant yield and the \dcat distribution and extract the invariant yield of charm and bottom hadron decay electrons in 0-10\% central Au+Au collisions. The inclusive invariant yield of electrons from heavy flavor hadron decays measured in 2004 is used to normalize the yield.

\section{Results}
\begin{figure}[htbp]
 \begin{minipage}{0.5\hsize}
  \begin{center}
   \includegraphics[width=1.0\columnwidth]{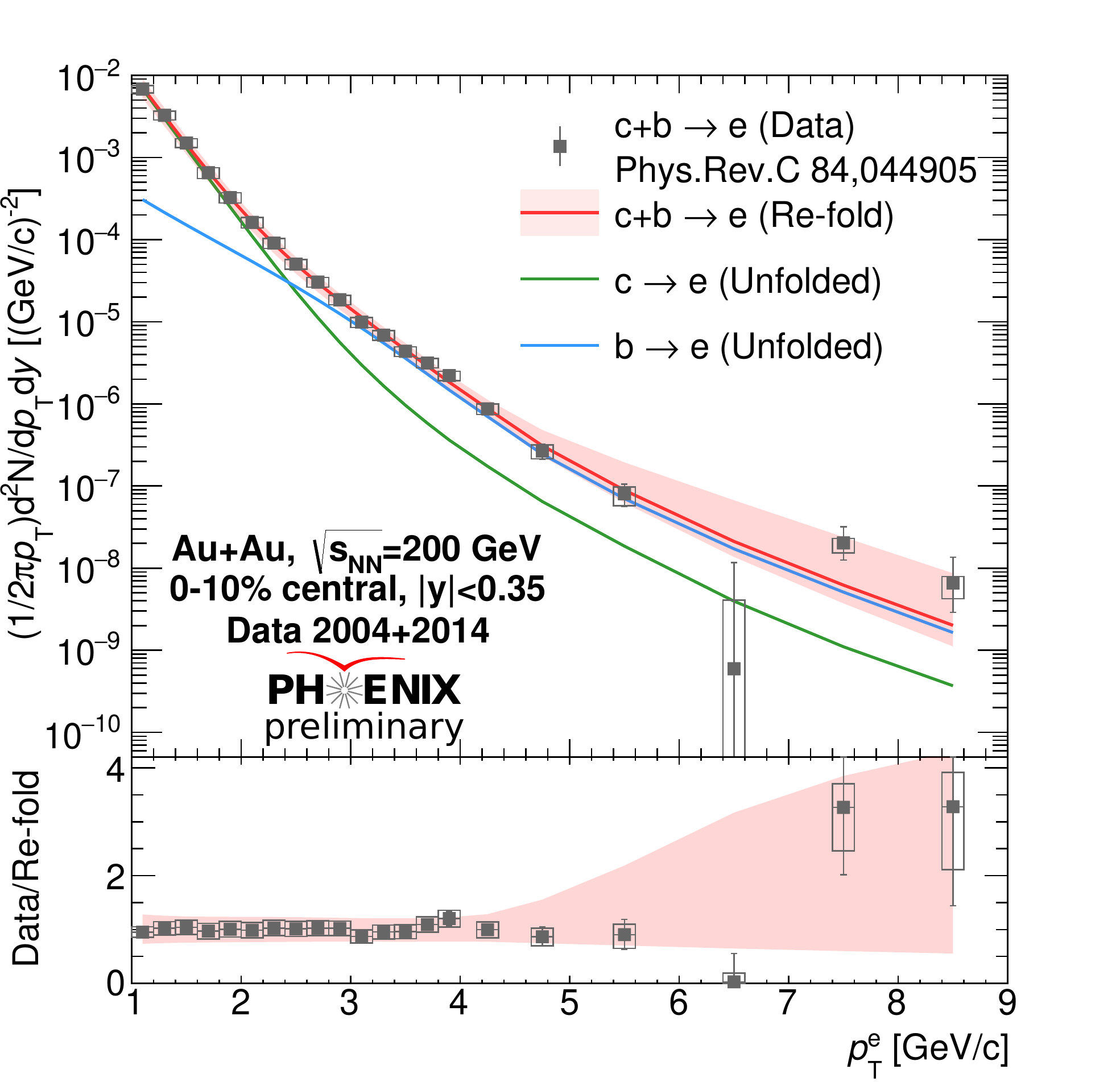}
  \end{center}
 \end{minipage}
 \begin{minipage}{0.5\hsize}
  \begin{center}
   \includegraphics[width=1.0\columnwidth]{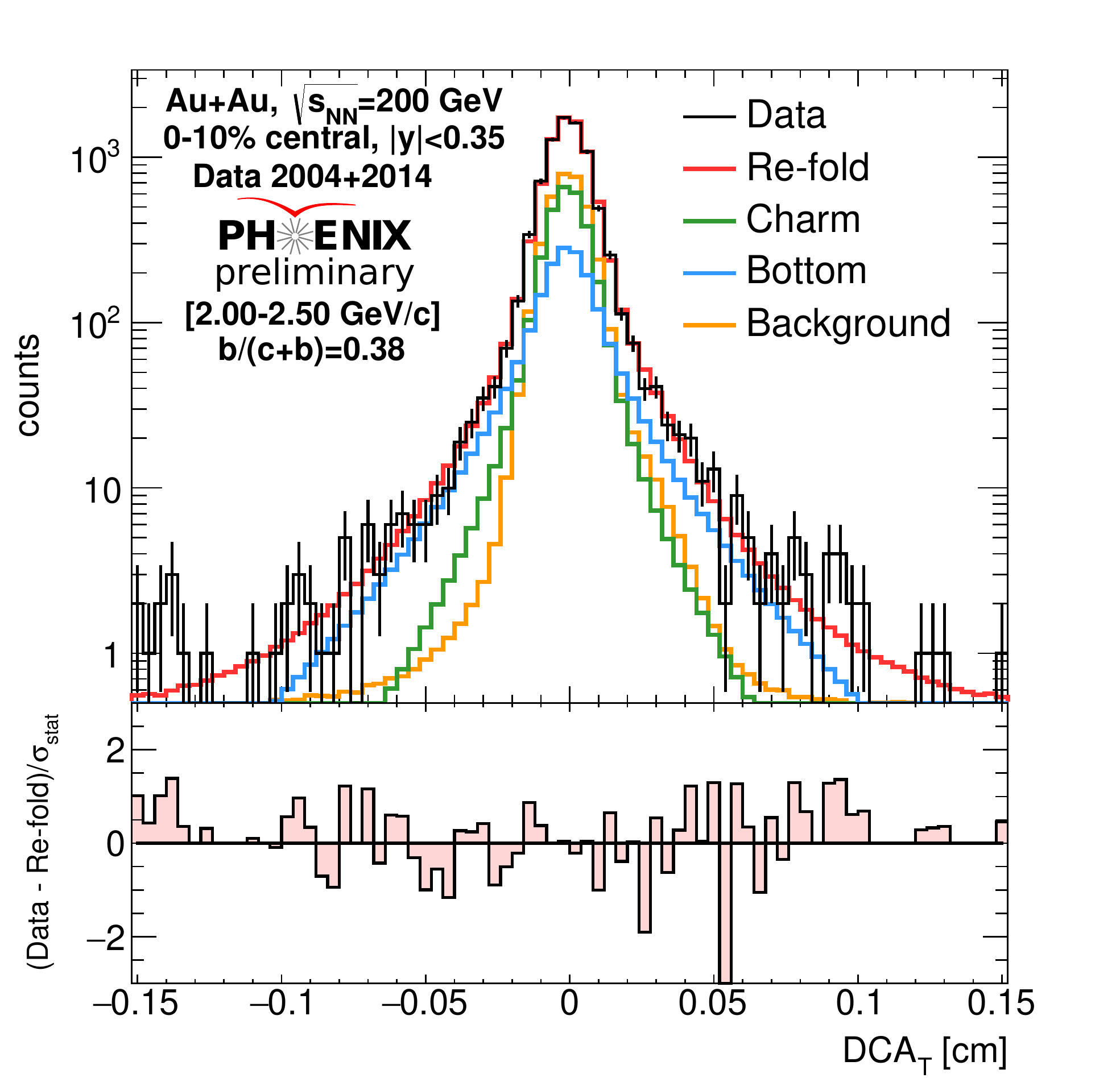}
  \end{center}
 \end{minipage}
 \caption{(The invariant yield (left) and the \dcat distribution (right) for measured electrons compared with unfolded charm (green line) and bottom (blue line) hadron decay electrons in 0-10\% central Au+Au collisions at \sqsn $=$ 200 \gevc. \label{fig:2}}
\end{figure}

\begin{figure}[htbp]
 \begin{minipage}{0.5\hsize}
  \begin{center}
   \includegraphics[width=1.0\columnwidth]{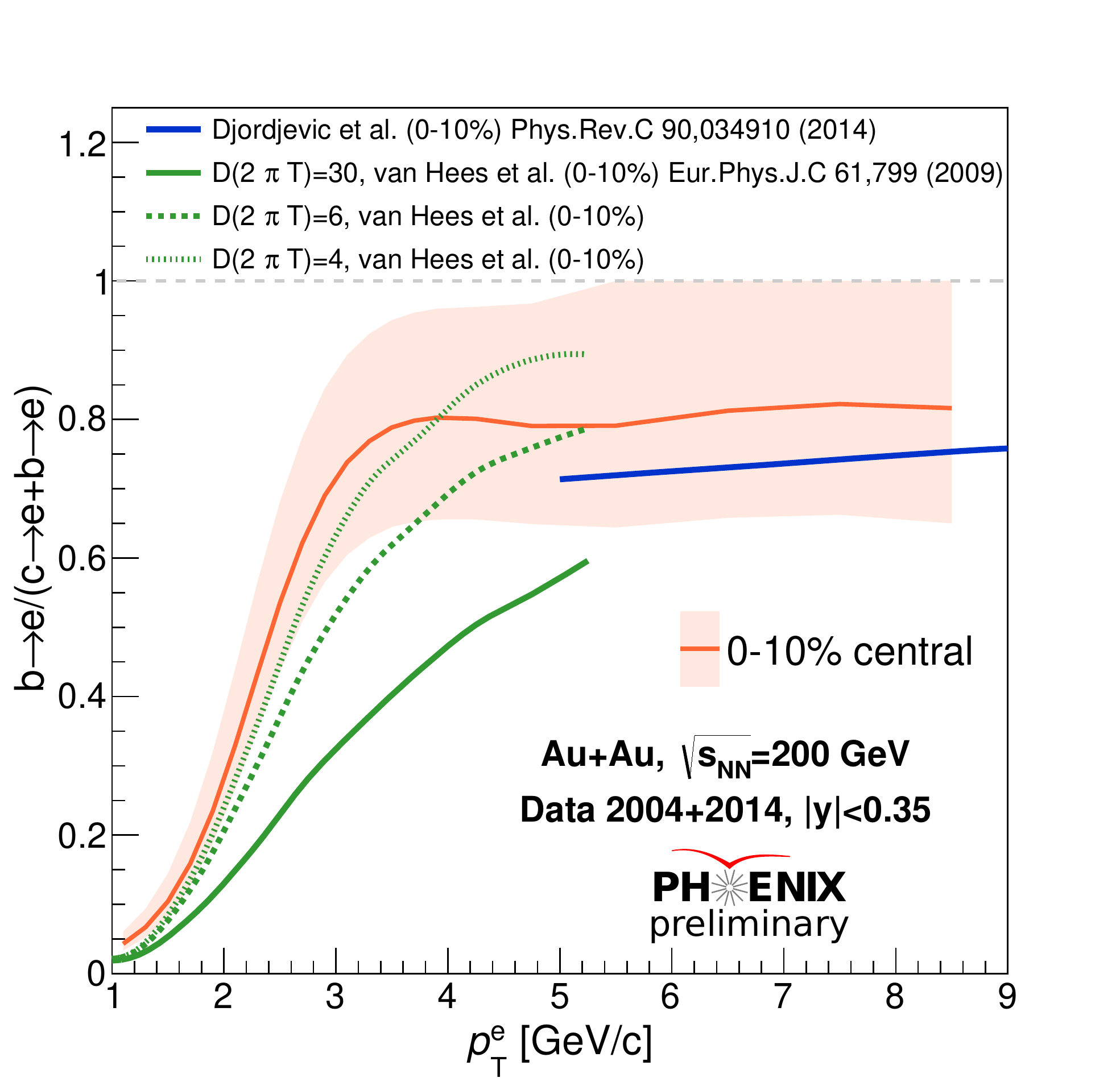}
  \end{center}
 \end{minipage}
 \begin{minipage}{0.5\hsize}
  \begin{center}
   \includegraphics[width=1.0\columnwidth]{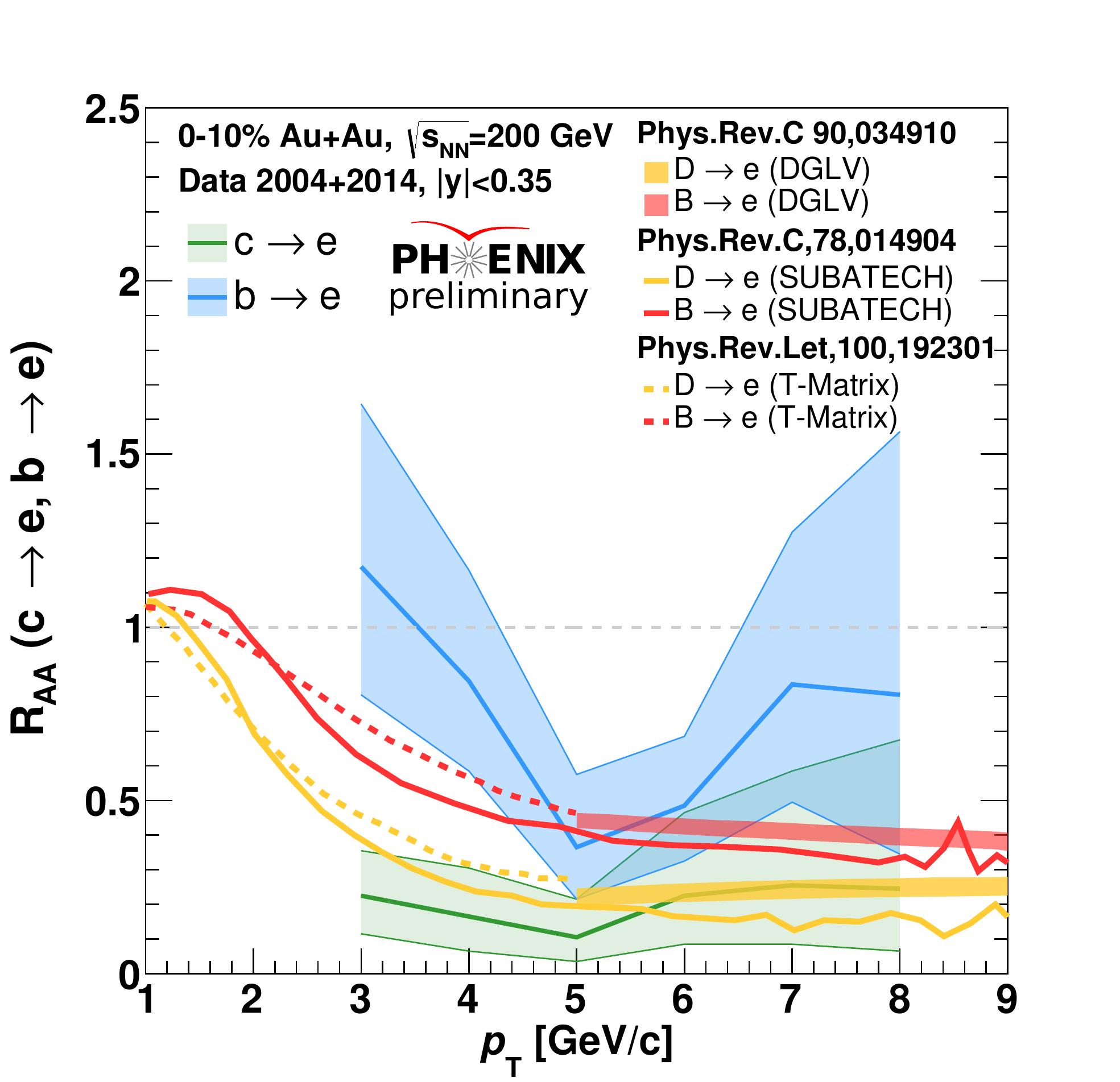}
  \end{center}
 \end{minipage}
 \caption{The bottom electron fraction (left) and the nuclear modification factor of electrons from charm (green line) and bottom (blue line) hadron decays in 0-10\% central Au+Au collisions, and theoretical model predictions. \label{fig:3}}
\end{figure}

 The invariant yields and \dcat distribution of separated charm and bottom hadron decay electrons are shown in Fig.\ \ref{fig:2}. The \dcat distribution provides a strong constraint of the bottom electron fraction (\btoe /(\ctoe+ \btoe)), shown in Fig.\ \ref{fig:3} (left), because of the different shapes of the \dcat distribution between charm and bottom hadron decay electrons. The nuclear modification factor, \raa, of separated charm and bottom hadron decay electrons are calculated using the inclusive \raa and the bottom electron fraction in \auau and \pp \cite{cite4} and shown in Fig.\ \ref{fig:3} (right).\par

 We find that electrons from bottom hadron decays are less suppressed than those from charm hadron decays in 3.0-5.0 \gevc in 0-10\% central Au+Au collisions. This result is in agreement with the theoretical prediction of the quark mass hierarchy of the energy loss in the QGP, ${\Delta}E_{\rm{c}} > {\Delta}E_{\rm{b}}$. The full statistics in 2014-2016 data will be needed to understand the quark mass hierarchy for \pt $>$ 5 \gevc because of the large statistical uncertainties. Note that the difference of \raa between charm and bottom hadron decay electrons is also included the difference of masses and fragmentation functions between charm and bottom quarks.\par
 We compare our extracted \raa to several theoretical models. We find that the T-Matrix calculation \cite{cite6} parametrized as $D(2\pi\it{T}) =$ 4 is in very good agreement with the measured bottom electron fraction in 0-10\% central Au+Au collisions. It favors that heavy quarks are strongly coupled in QGP. We also find that transport and energy loss models \cite{cite6,cite7,cite8} reproduce the measured \raa of both charm and bottom hadron decay electrons for \pt $>$ 4.5 \gevc. However, theoretical model predictions are slightly off the measured $\raa({\btoe})$ for \pt $<$ 4.5 \gevc. It seems that theoretical models may need stronger interaction in QGP to be more consistent with our data.

\section{Summary}
 PHENIX has measured the invariant yield of separated charm and bottom hadron decay electrons and the bottom electron fraction in 0-10\% central Au+Au collisions via applying the Bayesian inference techniques simultaneously to the inclusive invariant yield and \dcat distributions. The nuclear modification factor, \raa, of separated charm and bottom hadron decay electrons are extracted with the inclusive \raa and the bottom electron fraction in \auau and \pp.  We found that a quark mass dependence of \raa, $\raa(\ctoe) < \raa(\btoe)$ is clearly shown in 3.0-5.0 \gevc. It favors the quark mass hierarchy of the energy loss in the QGP, ${\Delta}E_{\rm{c}} > {\Delta}E_{\rm{b}}$. A T-Matrix model parametrized as $D(2\pi\it{T}) =$ 4  reproduces the measured bottom electron fraction in 0-10\% central Au+Au collisions and indicates that heavy quarks are strongly coupled in the QGP \cite{cite6}.  Current theoretical models \cite{cite6,cite7,cite8} also reproduce the measured $\raa(\ctoe)$ and $\raa(\btoe)$ for \pt $>$ 4.5 \gevc. However, they under-predicts are slightly off  $\raa(\btoe)$ for \pt $<$ 4.5.\par
  In the future, the full statistics collected in 2014-2016 should provide a high-precision measurement of $\raa(\ctoe)$ and $\raa(\btoe)$ for a broader \pt range, 1.0-9.0 \gevc, and 5 centrality bins. These will constrain models of quark energy loss, the quark mass hierarchy of the energy loss, and physical properties of the QGP.





\bibliographystyle{elsarticle-num}
\bibliography{<your-bib-database>}



\end{document}